\documentclass[fleqn,usenatbib,a4paper]{mnras}

\usepackage{newtxtext,newtxmath}
\usepackage[T1]{fontenc}
\usepackage{ae,aecompl}

\usepackage{graphicx}	% Including figure files
\usepackage{amsmath}	% Advanced maths commands
\usepackage{amssymb}	% Extra maths symbols

\newcommand{\Hmocks}{\textsc{hits-mocks}}
\newcommand{\Dmocks}{\textsc{icc-mocks}}

\newcommand{\gaia}{\textsc{gaia}}
\newcommand{\auriga}{\textsc{auriga}}
\newcommand{\aurigaia}{\textsc{aurigaia}}

\newcommand{\Msun}{M$_\odot$}

\newcommand\textlcsc[1]{\textsc{\MakeLowercase{#1}}}

\title[Phase space substructure in Auriga]{Simulating cosmological substructure in the solar neighbourhood}

\author[Simpson et al.]{\parbox[t]{\textwidth}{
Christine M. Simpson$^{1,2}$\thanks{csimpson@astro.uchicago.edu}, Ignacio Gargiulo$^{3,4}$, Facundo A. G\'{o}mez$^{3,4}$, Robert J. J. Grand$^{5}$, Nicol\'{a}s Maffione$^{6,7}$, Andrew P. Cooper$^8$, Alis J. Deason$^{9}$, Carlos Frenk$^9$, John Helly$^9$, Federico Marinacci$^{10}$, R\"{u}diger Pakmor$^5$ \vspace{10pt} }\\
$^1$Enrico Fermi Institute, The University of Chicago, Chicago, IL 60637, USA\\
$^2$Department of Astronomy \& Astrophysics, The University of Chicago, Chicago, IL 60637, USA\\
$^{3}$Instituto de Investigaci{\'o}n Multidisciplinar en Ciencia y Tecnolog{\'i}a, Universidad de La Serena, Ra{\'u}l Bitr{\'a}n 1305, La Serena, Chile\\
$^{4}$Departamento de F{\'i}sica y Astronom{\'i}a, Universidad de La Serena, Av. Juan Cisternas 1200 N, La Serena, Chile\\
$^5$Max-Planck-Institut f\"{u}r Astrophysik, Karl-Schwarzschild-Str. 1, D-85748, Garching, Germany  \\
$^6$Laboratorio de Procesamiento de Se\~{n}ales Aplicado y Computaci\'{o}n de Alto Rendimiento, Sede Andina, \\Universidad Nacional de R\'{i}o Negro, Mitre 630, San Carlos de Bariloche, R8400AHN R\'{i}o Negro, Argentina\\
$^7$CONICET, Mitre 630, San Carlos de Bariloche, R8400AHN R\'{i}o Negro, Argentina\\
$^8$Institute of Astronomy, Department of Physics, National Tsing Hua University, Hsinchu 30013, Taiwan\\
$^9$Institute for Computational Cosmology, Department of Physics, Durham University, South Road Durham DH1 3LE, UK \\
$^{10}$Department of Physics \& Astronomy, University of Bologna, via Gobetti 93/2, 40129 Bologna, Italy
}

\date{Accepted XXX. Received YYY; in original form ZZZ}

\pubyear{2018}

\begin{document}
\label{firstpage}
\pagerange{\pageref{firstpage}--\pageref{lastpage}}
\maketitle

\begin{abstract}
We explore the predictive power of cosmological, hydrodynamical simulations for stellar phase space substructure and velocity correlations with the \auriga\ simulations and \aurigaia\ mock-\gaia\ catalogues.  We show that at the solar circle the \auriga\ simulations commonly host phase space structures in the stellar component that have constant orbital energies and arise from accreted subhaloes.  These structures can persist for a few Gyrs, even after coherent streams in position space have been erased.  We also explore velocity two-point correlation functions and find this diagnostic is not deterministic for particular clustering patterns in phase space.  Finally, we explore these structure diagnostics with the \aurigaia\ catalogues and show that current catalogues have the ability to recover some structures in phase space but careful consideration is required to separate physical structures from numerical structures arising from catalogue generation methods.  
\end{abstract}


%%%%%%%%%%%%%%%%%%%%%%%%%%%%%%%%%%%%%%%%%%%%%%%%%%

%%%%%%%%%%%%%%%%% BODY OF PAPER %%%%%%%%%%%%%%%%%%

\section{Introduction}

The hierarchical assembly of galaxies, driven by the evolution of a cosmic web of dark matter, is a cornerstone prediction of the $\Lambda$CDM paradigm.  Observational evidence for this process has been found in the local Universe in the form of stellar streams and moving groups and has been the focus of studies empirically probing the assembly history of galaxies \citep{1999Natur.402...53H,Belokurov2006,Martinez-Delgado2010}.  Theoretical work in this area has focused on the dynamical evolution of this debris and used numerical simulations to make predictions for its connection to galaxy assembly histories \citep[e.g.][]{1999MNRAS.307..495H,BullockJohnston,2010MNRAS.406..744C,2010MNRAS.401.2285G,2015A&A...584A.120B,Maffione2018}.

A new era in this field has arrived, driven by the \gaia\ mission \citep{Gaia} that has identified many new streams and structures in the Galaxy \citep{2018Natur.561..360A,2018MNRAS.481.3442M,Helmi2018,Fragkoudi2019,2018ApJ...856L..26M}.  Cosmological simulations, the main tool for making theoretical predictions in the $\Lambda$CDM paradigm, have also made advances, and multiple simulation codes can now produce realistic stellar discs with baryon physics at high-resolutions \citep{Grand2017,hopkins2018}.  In this letter, we seek to highlight this advance and demonstrate the ability of simulations to produce realistic stellar phase space structures in solar neighbourhood-like volumes; we will focus on structures arising from cosmological accretion events.  These models can now be used as a theoretical platform to test ideas about galaxy evolution and dark matter structure.  Previous studies focusing on this volume have either considered models with an analytic galactic potential \citep{2010MNRAS.408..935G,2000MNRAS.319..657H} or dark matter-only cosmological simulations \citep[e.g.][]{2003MNRAS.339..834H,Vogelsberger2009,2013MNRAS.436.3602G}.

\begin{figure}
	\includegraphics[width=\columnwidth]{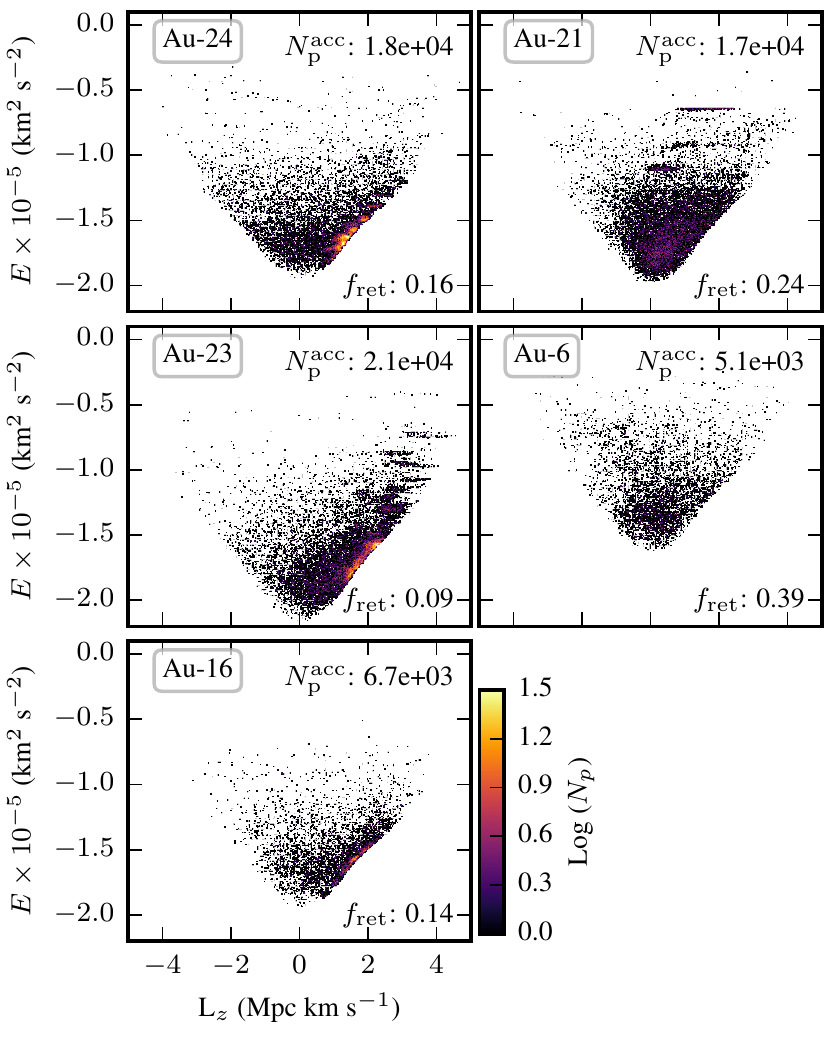}
    \caption{{\it E}-{\it $L_z$} space distributions of accreted star particles inside a 2.5 {\rm kpc}
    sphere positioned 8 {\rm kpc} from the galactic centre in the discs of five \auriga\ haloes at redshift 0. The number of particles in each panel is indicated ($N^{\textrm{acc}}_{\textrm{p}}$) and the fraction of the accreted material that is counter rotating ($f^{\textrm{}}_{\textrm{ret}}$). }
        \label{fig:L3_volumes}
\end{figure}

This letter is structured as follows: first, we describe the \auriga\ simulations and \aurigaia\ mock \gaia\ catalogues used for this study; second, we explore the typical properties of phase space structures at solar-like positions within \auriga; third, we consider mock observations, what aspects of phase space they are able to recover, and quantify their limitations; and finally, we conclude and discuss future directions for making predictions for stellar catalogues from simulations.

\section{The \auriga\ simulations and the \aurigaia\ mock stellar catalogues}

The \auriga\ simulations are a suite of cosmological, magneto-hydrodynamical zoom simulations of Milky-Way like disc galaxies \citep{Grand2017}.  The \auriga\ model includes many of the relevant physics for galaxy formation, including gravity, gas cooling, star formation, and stellar feedback effects in the form of energy and metal feedback.  A set of zoom haloes were selected in the mass range $1$-$2\times 10^{12}$ \Msun, giving a set of systems that have variations in their assembly histories, but yield similar final galaxies.  The zoom halo forms hierarchically from the accretion smaller haloes and hosts a population of surviving satellite dwarf galaxies at redshift zero \citep{Simpson2018}.  We study the six highest resolution Auriga haloes that have a typical star particle mass of $6\times 10^3$ \Msun\ and a gravitational force softening of 184 pc after $z=1$.  This high mass resolution enables exquisite sampling of the gravitational potential and the dynamics of stars, which is crucial to capture complex phase space structures like moving groups, streams and satellite debris. 

The \auriga\ simulations were used to create the \aurigaia\ catalogues, a suite of mock catalogues of the \gaia\ mission's Data Release 2 (DR2) \citep{Grand2018c}.  The \aurigaia\ catalogues contain `child' stars generated from `parent' star particles in the \auriga\ simulations.  Two sets of catalogues were created independently using different codes for four observer positions within each \auriga\ halo.  Both methods include the DR2 errors and selection function and empirical dust extinction from maps of the Milky Way's dust distribution.  The \Hmocks\ were created with a parallel version of the publicly available \textlcsc{SNAPDRAGONS} code\footnote{Available at \href{https://github.com/JASHunt/Snapdragons}{\url{https://github.com/JASHunt/Snapdragons}}} .
The \Dmocks\ were created following the method of \citet{Lowing2015}.  In addition to the effects in the \Hmocks, the \Dmocks\ distribute stars over a 6D kernel approximating the phase-space volume of their parent particles. 
We use the mock catalogues to understand how simulated phase space substructure is affected by the biases and errors of \gaia\ DR2 and by the methods of mock-catalogue creation.

\section{Accreted phase space structures in the Auriga simulations}     

We begin by looking at star particles in solar neighbourhood-like volumes in the \auriga\ simulations.
Star particles are selected within 2.5 kpc radius spheres, centred at a mid-plane position in the stellar disc (defined following \citet{Grand2017}) 8 kpc from the galaxy centre.  The \auriga\ discs vary in size and mass, and so 8 kpc corresponds to different normalized disc scale lengths between galaxies \citep{Grand2018c}.  The volumes discussed here have comparable mass surface density and circular velocity to the solar neighborhood, although the exponential scale lengths of the discs tend to be larger than the MW.  We do not find any of the structures discussed here correlate with this variance.  Particle specific energies and angular momenta are computed in a galactic rest frame aligned with the disc.  Fig.~\ref{fig:L3_volumes} shows accreted star particles in four example volumes.

The total specific energy of a particle is $E=0.5 v^2 + u$,
where $v$ is the magnitude of the particle's velocity vector $\mathbf{v}$ and $u$ is the specific potential energy of the particle computed from the mass density field truncated at the radius where the average density of the halo equals 200 times the critical density of the universe.  The specific angular momentum is $\mathbf{L} = \mathbf{r} \times \mathbf{v}$, where $\mathbf{r}$ is the particle's position vector.  The component of $\mathbf{L}$ along the disc rotational axis is $L_z$. 

The $E - L_z$ phase space is well suited to reveal signatures of accretion.  Indeed, satellite debris of common origin tends to clump in this space because $E$ and $L_z$ are close to true integrals of motion of the system \citep[and would be perfectly so for an axisymmetric system,][]{2000MNRAS.319..657H}. For the first time in a hydrodynamical cosmological simulation, we show clear signatures of stellar phase-space substructure in a solar neighbourhood-like volume.  Previous works in this area \citep[e.g.][]{2018arXiv180610564S} have mapped structures in position space on halo scales.  There is a diversity of structures from halo to halo and volume to volume, with some structures presenting a narrow range in $E$ and others showing larger spreads.  Some volumes have very little structure and others have substantial accreted discs.  The fraction of accreted material that is `counter-rotating' (i.e. $L_z < 0$) varies greatly ($10 - 40$ percent), however, the overall amount of counter-rotating material is small (less than 5 percent) due to the dominance of the in situ disc.

\begin{figure*}
	\includegraphics[width=\textwidth]{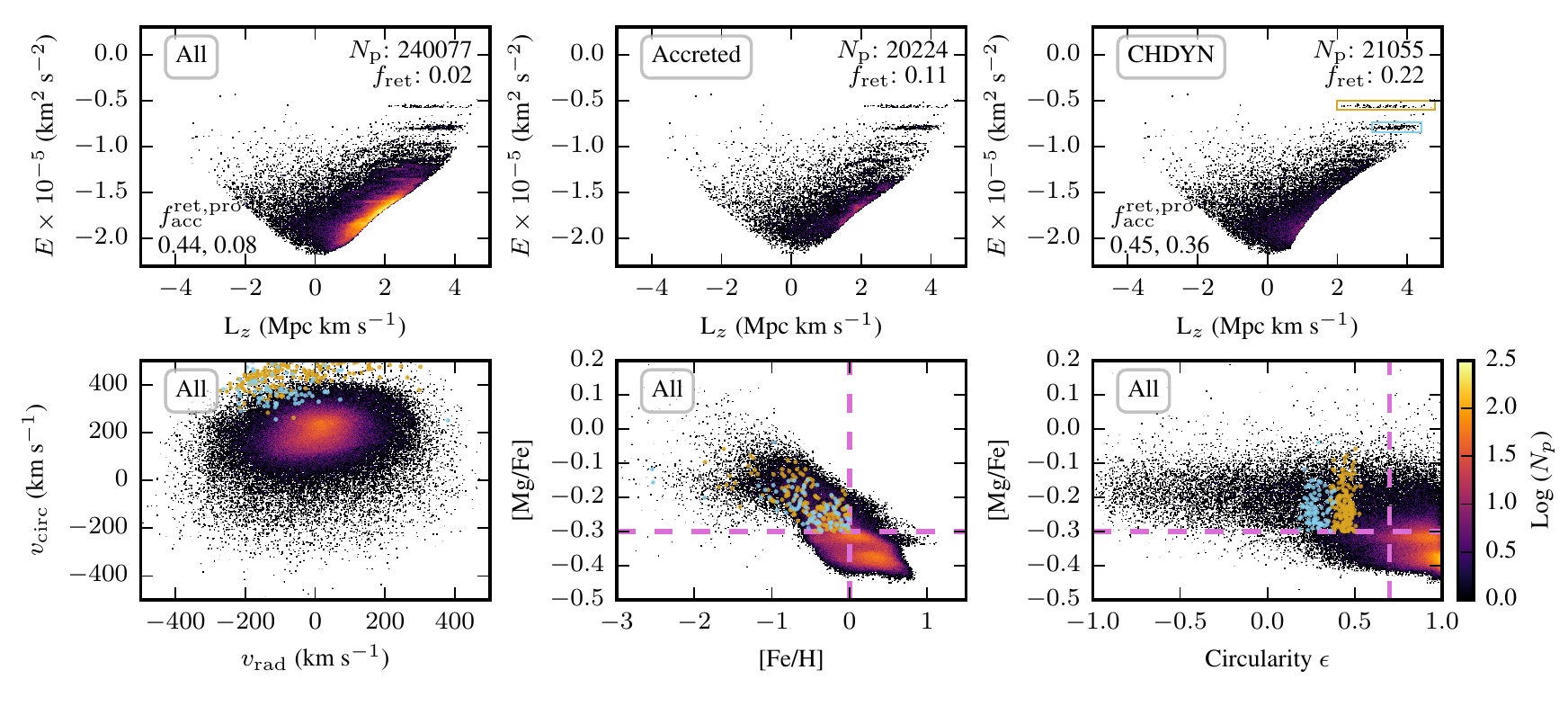}
    \caption{Top: $E$-$L_z$ phase space of star particles from Au-23.  Particles within a 2.5 kpc sphere centred 8 kpc from the disc centre are shown.  This volume is separated by 180 degrees from the Au-23 volume in Fig. \ref{fig:L3_volumes}.  All particles, accreted particles, and particles meeting a set of chemical and dynamical cuts described in the text (called the `CHDYN' sample) are shown separately. The number of particles ($N_\mathrm{p}$), the fraction of retrograde particles ($f_{\rm{ret}}$), and the fraction of prograde and retrograde particles that are accreted ($f^{\rm{ret,pro}}_{\rm{acc}}$) are indicated.  Two constant-energy accretion features at large $E$ are boxed in the CHDYN sample.  Bottom: Chemical and dynamical properties of particles in the volume.  Particles in the boxed accretion features are plotted with the corresponding colour from above.  Left: velocities in the direction of the disc rotation ($v_{\rm{circ}}$) vs. velocities in the radial direction ($v_{\rm{rad}}$).   Middle: [Mg/Fe] vs. [Fe/H].  Right: [Mg/Fe] vs. the orbital circularity $\epsilon$.  The CHDYN sample cuts are shown with dashed lines ([Fe/H] < 0, [Mg/Fe] > -0.3, and $\epsilon < 0.7$).}
    \label{fig:sim_phase_space}
\end{figure*}

Fig.~\ref{fig:sim_phase_space} shows a detailed look at one volume.  This volume has several structures apparent by eye as bands of constant $E$.  In position space, particles in these bands do not have any obvious structure.  In velocity space, they have a wide spread in radial velocity and a narrower spread in circular velocity.  These structures originated from the same accreted satellite that had several pericentric passages before destruction, which occurred 3 Gyrs before redshift 0.  The satellite had a peak stellar mass of $5 \times 10^9$ \Msun\ and its interaction with the host lasted for more than 3 Gyrs.  The two most prominent phase-space structures (boxed in Fig.~\ref{fig:sim_phase_space}) originated from the final pericentric passage and their energy difference is due to their association with the leading and trailing tidal debris of the interaction.  These structures have masses of $8.3 \times 10^5$ \Msun\ and $4.0 \times 10^5$ \Msun\ and the total mass of debris from their parent system in this volume is greater than $10^7$ \Msun.  

Overall, 8 percent of star particles in the volume shown in Fig.~\ref{fig:sim_phase_space} are accreted.  Fig.~\ref{fig:sim_phase_space} shows that many structures are accreted, including a disc component \citep{2017MNRAS.472.3722G}.  Some disc structures appear to be insitu, likely the result of secular disc dynamics.  Counter orbiting particles ($L_z < 0$) have a larger accreted fraction, 44 percent.  The overall fraction of counter rotating material, however, is small, only 2 percent.

Selecting stars by their chemical and orbital properties is a common technique used to eliminate in situ disc stars and isolate accreted and halo populations.  We experimented with this technique by using the abundances of H, Fe, and Mg tracked by the \auriga\ simulations.  We compute for each star particle an iron abundance relative to hydrogen and a magnesium abundance relative to iron in solar units ([Fe/H] and [Mg/Fe]).  Separating particles by these abundances probes material from stellar populations with different star formation histories (SFHs).  Dwarf satellite galaxies (the origin of cosmologically accreted stellar populations) with lower star formation rates and truncated or bursty SFHs will tend toward lower [Fe/H] and higher [Mg/Fe], due to the different timescales of iron production in Type Ia supernovae and $\alpha$-element production in Type II SN \citep{Tinsley79,2014MNRAS.444..515R}.  

Auriga captures the qualitative trend of [Mg/Fe] with [Fe/H] observed in the MW \citep{Grand2018}, however, the quantitative values for [Mg/Fe] are approximately 0.3 dex too low, an offset due to the choice of chemical yield tables.  We apply chemical cuts to the simulated data based on the qualitative trend.
We find 62 percent of accreted material in the volume has [Mg/Fe] > -0.3 and [Fe/H] < 0.  
There is also a significant population of in situ material with these abundances, so overall, only 28 percent of material meeting these cuts is accreted.  %

To isolate a halo population, we also apply a limit on the orbital circularity of star particles.  We determine as a function of $E$ the maximum $L_z$ available within the potential by examining the $L_z$ of all star particles within 50 kpc.  We define the orbital circularity to be $\epsilon_i = L_{z,i}/\mathrm{max(|}L_z(E_i)\mathrm{|)}$ \citep{2015MNRAS.454.3185C}.  We apply a limit of 0.7 in $\epsilon$ as shown in Fig.~ \ref{fig:sim_phase_space}.

We call the combination of these chemical and dynamical cuts the CHDYN sample ([Mg/Fe] > -0.3, [Fe/H] < 0 and $\epsilon <0.7$), shown in Fig.~ \ref{fig:sim_phase_space}.  It has a higher fraction of accreted material (38 percent versus 8 percent) and prominent phase space structures remain.  Table~\ref{tab} gives some properties of the CHDYN sample.  The accreted material in this volume is mostly part of the disc, so the $\epsilon$ cut means that the CHDYN sample contains just 39 percent of accreted particles.  The fraction of counter rotating material in this sample is larger than in the accreted sample, reflecting a contribution from the in situ stellar halo.  These cuts rely only on observable quantities, and we will therefore use them in our analysis of mock observations.

\begin{table}
\caption{Median statistics of solar spheres in \auriga.}
\label{tab}
\begin{tabular}{lccc}
\hline
Halo &Particle Number & Accreted    & Retrograde ($L_z < 0$)  \\
Number &($\times 10^4$) & fraction    & fraction \\
\hline
Au6 & 18.44 / 1.93 & 0.03 / 0.24 & 0.02 / 0.21\\
Au16 & 15.51 / 1.57 & 0.04 / 0.18 & 0.02 / 0.15\\
Au21 & 25.71 / 2.53 & 0.07 / 0.44 & 0.03 / 0.26\\
Au23 & 22.7 / 2.13 & 0.09 / 0.38 & 0.02 / 0.21\\
Au24 & 17.16 / 1.7 & 0.1 / 0.34 & 0.05 / 0.43\\
Au27 & 32.32 / 3.13 & 0.05 / 0.39 & 0.04 / 0.36\\
\hline     
\end{tabular}
\raggedright{Note: Quantities presented are the median of values for four 2.5  kpc radius spheres positioned 8 kpc from the disc centre and separated by 90 degrees.  The first value in each column is the median over all particles and the second is the median over particles in the CHDYN sample.}
\end{table}

\section{Mock observations}
\begin{figure}
	\includegraphics[width=\columnwidth]{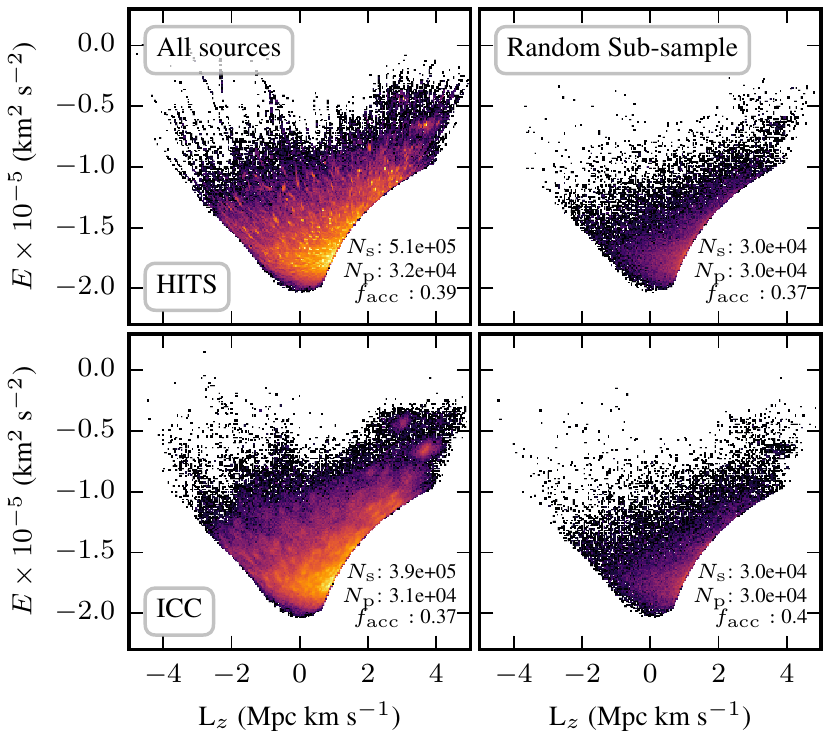}
	\includegraphics[width=\columnwidth]{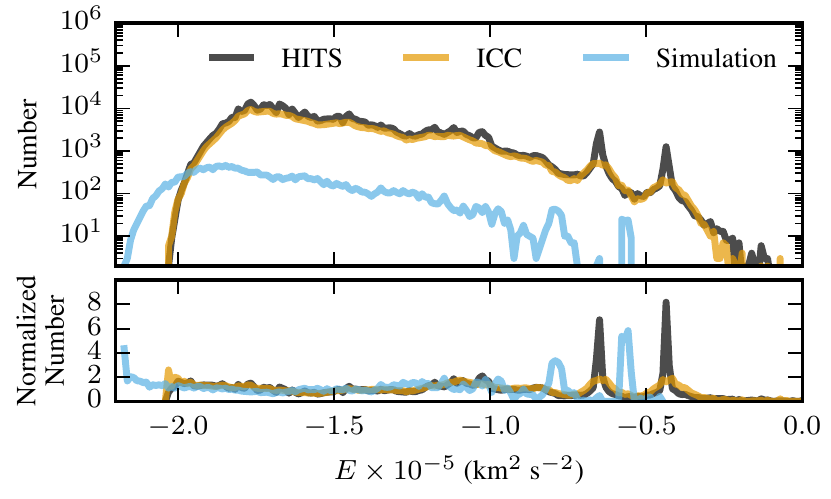}
    \caption{Top: $E$-$L_z$ phase space distributions of stars in the mock \gaia\ catalogues of the Au-23 volume shown in Fig. \ref{fig:sim_phase_space} for both the \Hmocks\ and \Dmocks.   CHDYN cuts are applied along with cuts particular to the mock catalogues (see text).  All stars meeting the selection cuts (left) and a random subsample of $3 \times 10^4$ stars with unique parent particles (right) are shown. The number of parent star particles ($N_p$), child stars ($N_s$), and the accreted fraction are indicated. Bottom: Distributions of $E$ for stars and star particles in the CHDYN sample in this volume in the mocks and simulation (top histogram) and normalized distributions (bottom histogram).  To normalize, we take the velocities of particles in the sample and randomize each velocity component independently.  $E$ is then computed with the new velocities.  This is repeated 100 times and an average $E$ distribution is computed from these trials.  The result is a distribution of $E$ that preserves the original distribution, but is free of particle-particle correlations.  The normalized distribution is the original distribution divided by this average, randomized distribution.}
    \label{fig:mock_phase_space}
\end{figure}

Fig. \ref{fig:mock_phase_space} shows the $E$-$L_z$ phase space in the \aurigaia\ mocks for the same volume shown in Fig.~\ref{fig:sim_phase_space} with the CHDYN sample selection cuts.  Potential energies for stars in the mocks are computed using a 3 component fit to the galactic potential following \citet{Maffione2018}.  For \aurigaia\ stars we assign the [Fe/H] and [$\alpha$/Fe] abundances of parent particles to all child stars.  We restrict analysis to stars that meet the radial velocity measurement criteria in DR2: $3 < G < 14$ and $3550 < T_{\rm{eff}} < 6900$, where $G$ is the the mean $G$ band magnitude and $T_{\rm{eff}}$ is the effective temperature.  We also require the error in parallax to be less than 10 percent.

The appearance of phase space between the two catalogues in Fig.~\ref{fig:mock_phase_space} is different despite the fact that they are generated from the same simulation, demonstrating the helpfulness of multiple approaches.  The \Dmocks\ utilise a phase-space smoothing technique that fills regions of phase space that are otherwise sparsely populated (while conserving the overall phase space density distribution).  The \Hmocks\ allow for discontinuities and gaps in phase space.  Both mocks show spurious structures, seeded by correlated sibling stars with the same parent particle.  Child stars' astrometric quantities are drawn from the same error distribution, resulting in correlated tracks in the $E-L_z$ plane.

To eliminate sibling star correlations, Fig.~\ref{fig:mock_phase_space} shows a random subsampling of stars from the mock catalogues with unique parent particles, i.e. no single parent particle contributes more than one child star to this sample.  We subsample at a number not exceeding the underlying resolution of the parent simulation.  Both mocks show clear overdensities associated with the phase space structures found in Fig. \ref{fig:sim_phase_space}, but that are stretched diagonally in this plane.  The overdensities are greatly reduced in the subsampled space, but are still visible by eye, especially in the \Dmocks.  

The structures are visible in distributions of energy shown in Fig.~\ref{fig:mock_phase_space}.   The number of mock stars at the energies of the structures exceed the background by a factor of 2 (\Dmocks) to 10 (\Hmocks).  The true over-density in the simulation data is 4-6.  There is an energy offset between the simulations and mocks due to an imperfect fit of the analytic gravitational potential.  The broadening in energy of the structures is primarily due to the displacement in velocity applied to catalogue stars from the \gaia\ errors and the phase space smoothing algorithm in the \Dmocks.  For an applied velocity difference $\mathbf{\varepsilon_v}$, the kinetic energy of a particle will be displaced by $\mathbf{\varepsilon_v} \cdot \mathbf{v} + 0.5\mathbf{\varepsilon_v}^2$. 

We also characterise the degree of coherent motions with a 2-point velocity correlation function (VCF).  Following \citet{Helmi2017}, we define the 2-point correlation function $\xi(\Delta v) = DD/\langle RR \rangle - 1$,
where $\xi$ is a function of $\Delta v$, which is the magnitude of the velocity difference of a pair of particles (i.e. $|\boldsymbol{v}_i - \boldsymbol{v}_j|$), $DD$ is the number of particles that have a velocity difference of $\Delta v$, and $\langle RR \rangle$ is the average number of random pairs with velocity difference $\Delta v$.  The average $\langle RR \rangle$ is computed by randomising two components of the velocity vectors of particles in the subsample being considered and computing the number of pairs in $\Delta v$.  This procedure is repeated 10 times and $\langle RR \rangle$ is the average value over these iterations.

\begin{figure}
	\includegraphics[width=\columnwidth]{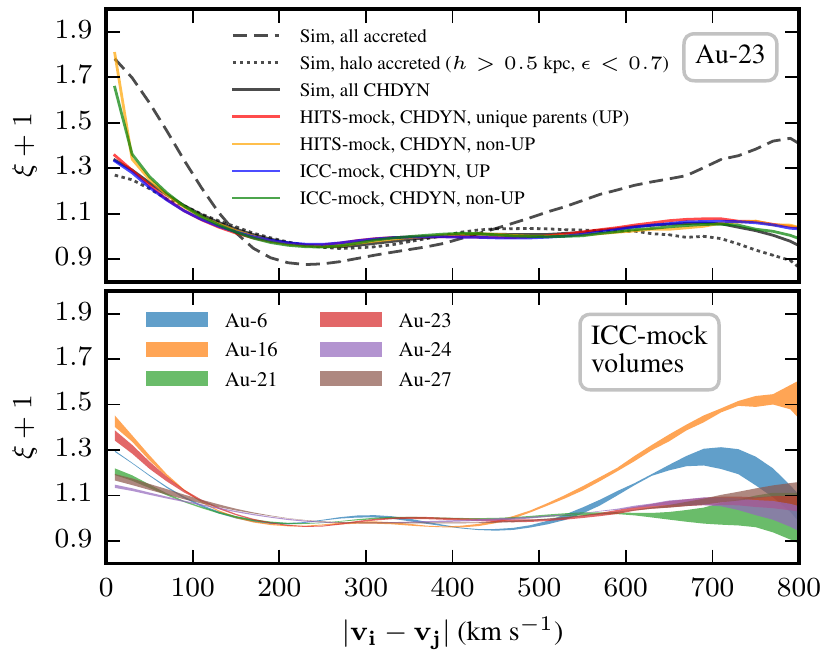}
    \caption{Top: VCFs of simulated and mock catalogue data for the Au-23 volume shown in Figs. \ref{fig:sim_phase_space}  and  \ref{fig:mock_phase_space}.  Curves for the accreted simulation sample and the CHDYN simulation and mock samples are shown.  An accreted halo simulation subsample is also shown that includes accreted particles in the volume with $\epsilon < 0.7$ more than 0.5 kpc above the disc. The mock data are subsampled by $3 \times 10^4$ as in Fig. \ref{fig:mock_phase_space}.  Mock subsampling is done in two ways: one sample randomly selects from all stars (non-UP) and one ensures unique parents (UP).  Bottom: VCFs for the \Dmocks\ catalogues created for all the haloes.  A number of CHDYN selected stars are randomly chosen with unique parents; the number selected is fixed to the number of unique parents contributing to the sample.  The range of results among the 4 curves is shown for each simulation as a shaded region. }
    \label{fig:twopoint}
\end{figure}

Computing VCFs with the mock data presents some of the same challenges as the phase-space maps.  Fig.~\ref{fig:twopoint} demonstrates that sibling stars can create a spurious low velocity difference excess. However, by restricting the mock sample to unique parent stars, Fig.~\ref{fig:twopoint} shows that it is possible to recover the same VCFs as seen in the true simulation data and achieve a result largely insensitive to the mock generation method.

The ultimate utility of VCFs is uncertain.  Previous work has interpreted excess at both high and low velocity differences as indicative of the dynamical state of accreted material \citep{Helmi2017}.  The CHDYN selection cuts were chosen to select a halo sample with a higher fraction of accreted stars.  The pure accreted sample has a large VCF excess at large and small velocity differences due to a massive accreted disc.    
Low velocity difference excess does not always correlate with the types of structures shown in Figs. \ref{fig:L3_volumes} and \ref{fig:sim_phase_space}, e.g. Au-23 has many structures and has a low velocity difference excess, but Au-16 does not and also has a high excess. In the case of Au-16, this excess appears to arise from contamination from the insitu disk; the CHDYN sample in Au-16 has an accreted fraction of only 18\%.  Table \ref{tab} shows that high velocity difference excess inversely correlates with the counter rotating fraction, however, this may be due to the low number of haloes.

\section{Discussion and conclusions}

We have explored phase space and velocity correlations in both simulations and mock stellar catalogues.  We find that cosmological simulations with baryon physics can produce stellar phase space structures from accretion events that persist for many Gyrs.  We have also shown that current state of the art mock catalogues can be used to detect and interpret simulated phase space structures in the observational plane, but care is needed to account for correlations between stars generated from the same simulation particle.  VCFs can be recovered, but the underlying physical reasons for features in VCFs are non-unique. 
By demonstrating the quantitative dimensions of these issues, this letter points toward numerical paths for improving observational predictions from simulations.  
For example, resolution for this use case is needed primarily in the gravitational dynamics of star particles; current state of the art simulations tie this resolution to the gas resolution, but this need not be the case.

This work also hints at interesting physics captured in the \auriga\ discs for future study.  The pattern in circular and radial velocities shown in Fig.~\ref{fig:sim_phase_space} is reminiscent of the observed velocity pattern \citep{Fragkoudi2019} perhaps indicating the influence of the Galactic bar or of a minor merger \citep{2012MNRAS.419.2163G}.  Prominent phase-space structures are dependent on the volume considered and may incompletely sample the dynamics of the full debris field \citep{2010MNRAS.401.2285G}; this can result in differing Galactic mass estimates from high-velocity stars \citep{Deason2019,Grandinprep}.  

The \gaia\ mission has expanded our understanding of the MW's accretion history.  Cosmological simulations will be an important tool for interpreting these data and this letter has provided the first proof of concept of this strategy.  We have described notable caveats to working with mock data, but we anticipate further improvements to simulations and mock generation methods will result in a robust tool for theoretical interpretation of the mission's data in this regime. 

\section*{Acknowledgements}

We thank the anonymous reviewer for their helpful comments that improved this letter.  FAG receives financial support from CONICYT through the project FONDECYT Regular Nr. 1181264. FAG and IG recieve financial support from the Max Planck Society through a Partner Group grant.  APC is supported by the Taiwan Ministry of Education Yushan Fellowship.  JH and CSF were supported by the Science and Technology Facilities Council (STFC) [grant number ST/F001166/1, ST/I00162X/1, ST/P000541/1]. CSF acknowledges European Research Council (ERC) Advanced Investigator grant DMIDAS (GA 786910). This work used the DiRAC Data Centric system at Durham University, operated by the ICC on behalf of the STFC DiRAC HPC Facility (www.dirac.ac.uk). This equipment was funded by BIS National E-infrastructure capital grant ST/K00042X/1, STFC capital grant ST/H008519/1, and STFC DiRAC Operations grant ST/K003267/1 and Durham University. DiRAC is part of the National E-Infrastructure.  FM acknowledges support through the Program ``Rita Levi Montalcini'' of the Italian MIUR.

%%%%%%%%%%%%%%%%%%%%%%%%%%%%%%%%%%%%%%%%%%%%%%%%%%



\bibliographystyle{mnras}
\bibliography{references} % if your bibtex file is called example.bib

% Don't change these lines
\bsp	% typesetting comment
\label{lastpage}
\end{document}